\documentclass[prl,twocolumn,english,showpacs]{revtex4-1}
\usepackage{graphicx}
\usepackage{amssymb}
\usepackage[rightcaption]{sidecap}
\usepackage{amsmath}
\usepackage{braket}
\usepackage[utf8]{inputenc}
\usepackage{hyperref}
\hypersetup{
    bookmarks=true,         
    unicode=false,          
    pdftoolbar=true,        
    pdfmenubar=true,        
    pdffitwindow=true,      
    pdftitle={Directed photon emission and enhanced absorption of a single ion coupled to a fiber-cavity},    
    pdfauthor={AQO group},     
    pdfsubject={single ion coupled to fiber cavity},   
    pdfnewwindow=true,      
    pdfkeywords={ion trap, fiber cavity, photons}, 
    colorlinks=true,       
    linkcolor=blue,          
    citecolor=red,        
    filecolor=black,      
    urlcolor=blue,           
}

\begin{document}

\title{Photon emission and absorption of a single ion coupled to an optical fiber-cavity}


\author{ M. Steiner$^{1,\dag}$, H. M. Meyer$^{1,2,\ast}$ J. Reichel$^{3}$,  M. K{\"o}hl$^{1,2}$\\
\normalsize{$^{1}$Cavendish Laboratory, University of Cambridge, JJ Thomson Avenue, Cambridge CB3 0HE, United Kingdom}\\
\normalsize{${^2}$Physikalisches Institut, University of Bonn, Wegelerstrasse 8, 53115 Bonn, Germany}\\
\normalsize{$^3$ Laboratoire Kastler Brossel, \'Ecole Normale Sup\'erieure,}\\
\normalsize{Universit\'e Pierre et Marie Curie–Paris 6, CNRS, 24 Rue Lhomond, 75005 Paris, France}\\
\normalsize{$^{\dag}$Present address: Centre for Quantum Technologies,
National University of Singapore,}\\ \normalsize{3 Science Drive 2; Singapore 117543; Singapore}\\
\normalsize{$^\ast$To whom correspondence should be addressed; E-mail:  hmeyer@physik.uni-bonn.de}
}

\begin{abstract}
We present a light-matter interface which consists of a single $^{174}$Yb$^+$ ion coupled to an optical fiber-cavity. We observe that photons at 935\,nm are mainly emitted into the cavity mode and that correlations between the polarization of the photon and the spin state of the ion are preserved despite the intrinsic coupling into a single-mode fiber. Complementary, when a faint coherent light field is injected into the cavity mode we find enhanced and polarization dependent absorption by the ion.
\end{abstract}


\date{\today}

\maketitle
Large-scale quantum networks will pave the way towards secure, long-distance communication via quantum cryptography and distributed quantum information processing~\cite{kimbleNature2008,PhysRevLett.75.3788,Duan2001,PhysRevLett.78.3221}. Crucial for the efficiency and feasibility of a quantum network is the interface which transforms between stationary and flying quantum bits. The most advanced stationary quantum bits, trapped atomic ions, have proven to be an ideal experimental platform to create, store and process quantum information, however, efficient conversion of quantum information back and forth to transmittable photonic quantum bits remains a challenge. 

Near-unity efficiency of photon emission and absorption can be achieved by coupling the stationary quantum bit to an optical cavity. In cavity quantum-electrodynamics (cavity-QED) the light--matter interaction can be dominated by the coherent coupling to a single mode of the electromagnetic field, defined by the cavity geometry. The coupling strength $g_0$ between a single emitter and a single cavity-photon depends on the electric dipole moment $d$ of the transition and the mode volume $V$ of the cavity, $g_0 \propto d/\sqrt{V}$. Albeit coherent control over the emission process has been demonstrated in ion-cavity systems with large mode volumes~\cite{Keller2004,PhysRevA.85.062308,Stute2012,StuteA.2013}, it is desirable to reduce the size of the cavity mode to increase the interaction strength. However, the miniaturization of the cavity is complicated by the high sensitivity of ions to dielectric surfaces~\cite{Harlander:2010}. 

In order to progress towards the strong coupling regime, we pursue the integration of fiber-based Fabry-P\'{e}rot cavities into radio-frequency (RF) Paul traps~\cite{PhysRevLett.110.043003,brandstaetter2013,Takahashi2013}. The key elements of the apparatus are the cavity mirrors which are micro-fabricated on the front facet of single-mode optical fibers~\cite{Hunger:2010}. Trapping ions within small optical cavities requires ion traps specifically designed to weaken the adverse effects of the dielectric surfaces on the trapping performance. Crucially, the small size of the fiber mirrors facilitates the incorporation in these microscopic traps. However, new questions arise about the effect of the intrinsic fiber coupling on faithful quantum state transfer to and from the ion because birefringence and mismatch between fiber and cavity mode potentially outweigh the benefits of these miniature cavities.

\begin{figure*}
\includegraphics[width=0.98\textwidth,angle=0]{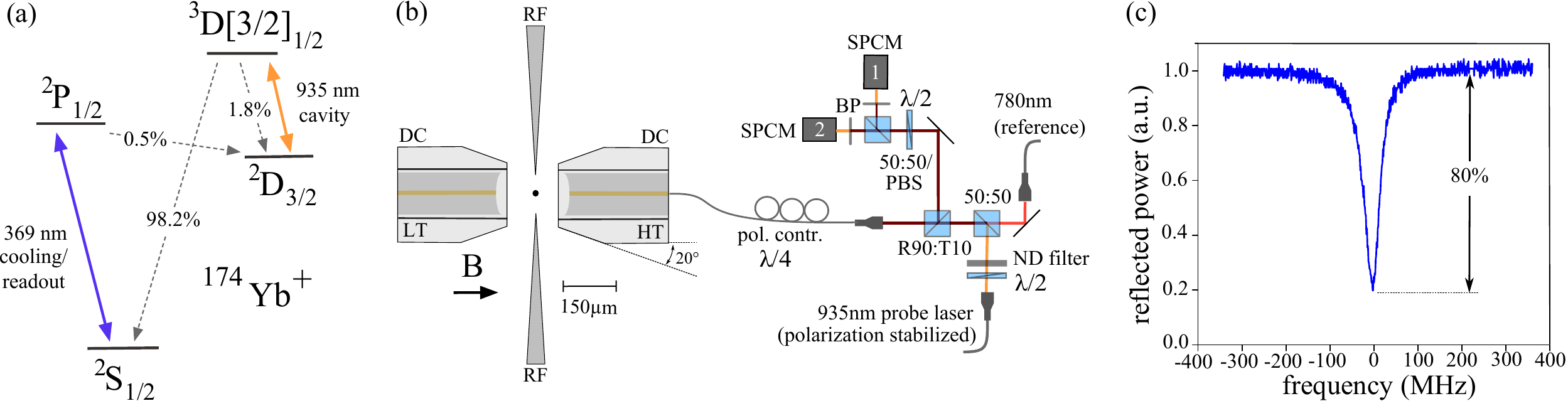}
\caption{\label{fig:fig_1}  (color online). \textbf{(a)}  Relevant level scheme and natural branching ratios of $^{174}$Yb$^+$ (not to scale). The cavity operates at the 935\,nm repump transition. \textbf{(b)}  Sketch of the ion trap with integrated fiber-cavity. An RF-signal of $100\,$V amplitude and $\Omega_\text{RF}/2\pi = 22\,$MHz frequency is applied to two needle-shaped electrodes separated by $100\,\mu \text{m}$ distance. The cavity fibers are placed inside stainless steel sleeves which are chamfered by an angle of $20^\circ$ to increase the optical access. The transmittance of the cavity mirrors is $100\,\text{ppm}$ at the high transmittance (HT) side and $10\,\text{ppm}$ at the low transmittance (LT) side. Optical setup: Photons from the HT-side of the fiber-cavity are guided onto two single-photon counting modules (SPCM). For the spin-photon correlation measurements the non-polarizing beam splitter (50:50) in front of the detectors is replaced by a polarizing beam splitter (PBS). The optical fiber attached to the HT-side of the cavity is adjusted to act as a quarter waveplate by applying mechanical stress to the fiber. The cavity length is actively stabilized by a piezoelectric transducer to a reference laser at 780\,nm. Abbreviations: BP: band pass at 935\,nm, $\lambda/2$: half wave-plate, R90:T10: non-polarizing beam splitter with 90\% reflection and 10\% transmission. \textbf{(c)} Coupling efficiency $\epsilon$ between cavity and the fiber mode at the HT-mirror. The reflection of a probe beam impinging upon the empty cavity drops on resonance by $80\,(5)\%$. Taking into account the finite in-coupling for perfect mode-matching we deduce $\epsilon = 90\,(5)\%$ (see supplementary materials).    }
\end{figure*}

Here, we address these issues by studying the photon emission and absorption properties of a single $^{174}\text{Yb}^+$ ion coupled to an optical fiber-cavity. The single ion is trapped in an RF-Paul trap with an ion-electrode distance of $50\,\mu$m  by applying a RF-signal (amplitude $100\,$V, frequency $22\,$MHz) to two opposing needle-shaped electrodes~\cite{PhysRevLett.110.043003,PhysRevLett.97.103007}. Laser cooling and fluorescence detection of the ion are performed on the $^2$S$_{1/2}$-$^2$P$_{1/2}$ transition near 369\,nm (Figure~1a). Spontaneous decay from the excited $^2$P$_{1/2}$ state populates the metastable $^2$D$_{3/2}$ state, which is cleared out by a repumping laser at 935\,nm.

The $170\,\mu$m-long high-finesse optical cavity is composed of two single-mode optical fiber ends and is placed perpendicular to the RF-electrodes (Figure~1b). Accumulation of laser-induced stray charges on the dielectric surfaces is minimized by mounting the cavity fibers inside of metal sleeves which protect them against laser radiation. The sleeves are held at a static electrical potential with respect to the radio-frequency signal. The cavity length is actively stabilized to be resonant with the $^2$D$_{3/2}$-$^3$D[$3/2$]$_{1/2}$ transition of  $^{174}\text{Yb}^+$ at a wavelength $\lambda=935\,$nm. With a finesse of $2.0\,(2)\times10^4$ the cavity field decay rate is $\kappa=2\pi \times 25\,(5)\,\text{MHz}$, improving an order of magnitude over our previous work~\cite{PhysRevLett.110.043003}. The mirrors have a radius of curvature of $300\,(50)\,\mu$m, which leads to a waist of $w_0=6.1\,\mu$m for the fundamental TEM$_{00}$ mode and a corresponding mode volume of $6000\lambda^3$. The transmittance of the cavity mirrors is asymmetric with a ratio of 1:10, hence photons are mainly transmitted through the high transmittance mirror (HT-mirror). The out-coupling efficiency for a cavity photon is limited by passive cavity loss to $\eta_\text{HT-mirror}=32(4)\%$  (see supplementary  materials). The experimentally observed cavity-QED parameters read ($\bar{g}_0$,~$\kappa$,~$\gamma$)$=2\pi\times$($1.6$,~$25$,~$2.11$)~MHz, where $\bar{g}_0$ is the coupling strength for the transition with the largest Clebsch-Gordan coefficient and $\gamma$ is the dipole decay rate of the $^3$D[$3/2$]$_{1/2}$ state. This places the cavity in the intermediate coupling regime of $\bar{g}_{0}\approx\gamma < \kappa$.


One major benefit of the cavity-based light-matter interface is that fluorescence can be efficiently extracted because the photon emission is channeled into a single mode \cite{McKeever26032004,Keller2004,Hijlkema2007,1367-2630-11-10-103004}. The experimental sequence to probe this behavior and to produce photons is as follows: the ion is manipulated by laser beams transverse to the cavity axis, starting with the preparation into the $^2$D$_{3/2}$ state by optical pumping on the 369\,nm transition. Subsequently a laser pulse at 935\,nm excites the ion resonantly into the $^3$D[$3/2$]$_{1/2}$ state from where it decays either to the $^2$S$_{1/2}$ state or back to the $^2$D$_{3/2}$ state. In free space the decay to the $^2$S$_{1/2}$ state is dominant with a probability of 98.2\% \cite{0953-4075-31-15-006}. However, the presence of the cavity modifies the atomic branching ratio, and in the intermediate coupling regime the probability for photon emission into the cavity mode $p_\text{emit}$ is given by the single atom cooperativity $ C_0= g_0^2/(2\kappa \gamma)$~\cite{citeulike:7719062}
\begin{equation}
p_\text{emit}=\frac{2C_0}{2C_0+1}~.
\end{equation} 
Given an excitation of the ion into the $^3$D[$3/2$]$_{1/2}$ state, the total probability $\eta_\text{total}$ to detect a photon at the HT-side of the cavity is 
\begin{equation}\label{eq:total_eta}
\eta_\text{total} =  p_\text{emit} \cdot  \underbrace{\eta_\text{HT-mirror}}_{32(4)\%} \cdot  \underbrace{\epsilon}_{90(5)\%} \cdot  \underbrace{\eta_\text{path}}_{75(5)\%} \cdot  \underbrace{\eta_\text{det}}_{25\%}
\end{equation}
where $\eta_\text{HT-mirror}$ is the probability for transmission through the HT-mirror, $\epsilon$ is the mode overlap between cavity and the single mode fiber (Figure~1c), $\eta_\text{path}$ is the attenuation due to optical path loss and  $\eta_\text{det}$ is the efficiency of the employed single-photon counting modules (SPCM). 
For each excitation into the $^3$D[$3/2$]$_{1/2}$ state we measure a photon detection probability of $\eta_\text{total, exp}= 0.33(4)\%$ after correcting for background light and multiple excitations. Thus, from $\eta_\text{total, exp}$ and Eq.~\ref{eq:total_eta} we deduce $p_\text{emit} = 6.1(1.5)\%$, $ C_0=0.032(8)$  and $\bar{g}_0=2\pi\times 1.6(2)\,$MHz (see supplementary  materials). Although the cavity-ion system does not operate in the strong coupling regime ($g_0>\kappa,\gamma$), the probability to emit a photon into the single cavity mode is already three times higher than the natural 935\,nm decay into all free space modes. Another important figure of merit for a photon collection system is the total probability to produce a photon into a single-mode fiber 
\begin{equation}
p_\text{emit,fiber}=p_\text{emit} \cdot \eta_\text{HT-mirror} \cdot \epsilon =1.8 (5) \%~.
\end{equation}

Next, we analyze the time-arrival histogram and the temporal photon statistics to demonstrate the generation of single photons with a spectral bandwidth that is lifetime limited. To this end, we employ $2.7\,$ns-long resonant excitation pulses at 935\,nm subsequent to the preparation of the ion in the $^2$D$_{3/2}$ state. The pulse duration is an order of magnitude shorter than the excited state lifetime $\tau_{[3/2]}=37.7\,\text{ns}$, hence multiple excitations are negligible. The time-arrival histogram (Figure~2a) shows a sharp rise due to the near-instantaneous excitation followed by a slow decay. The interaction with the cavity field causes the intensity of the ion emission to decay faster ($\tau_\text{hist}=35(1)\,$ns) than the natural excited state lifetime $\tau_{[3/2]}$. Using the cavity-QED parameters and taking account of background light, the temporal shape is well reproduced by the analytical expression for instantaneous excitation~\cite{PhysRevA.73.053807} as well as the numerical solution of the master equation.  


In a Hanbury-Brown and Twiss experiment we measure the second order correlation function and verify that the emitted light consists of a stream of single photons. The excitation cycle is repeated every $4\,\mu$s and the detection events during a 130\,ns interval shortly after the excitation pulse are integrated. The histogram of the coincidence detection events of detector 1 and 2 separated by a time $\tau$ shows a reduced probability to find detection events on both detectors during the same excitation cycle $\tau=0$ (Figure~2b). The photon anti-bunching confirms the non-classical nature of the emitted light. The detection interval does not cover the first 20\,ns after the excitation pulse because of background light caused by the resonant excitation pulse. The number of residual coincidence events at $\tau=0$ is expected from the background and dark count rate of the SPCMs.
\begin{figure}[!h]
\includegraphics[scale=0.42]{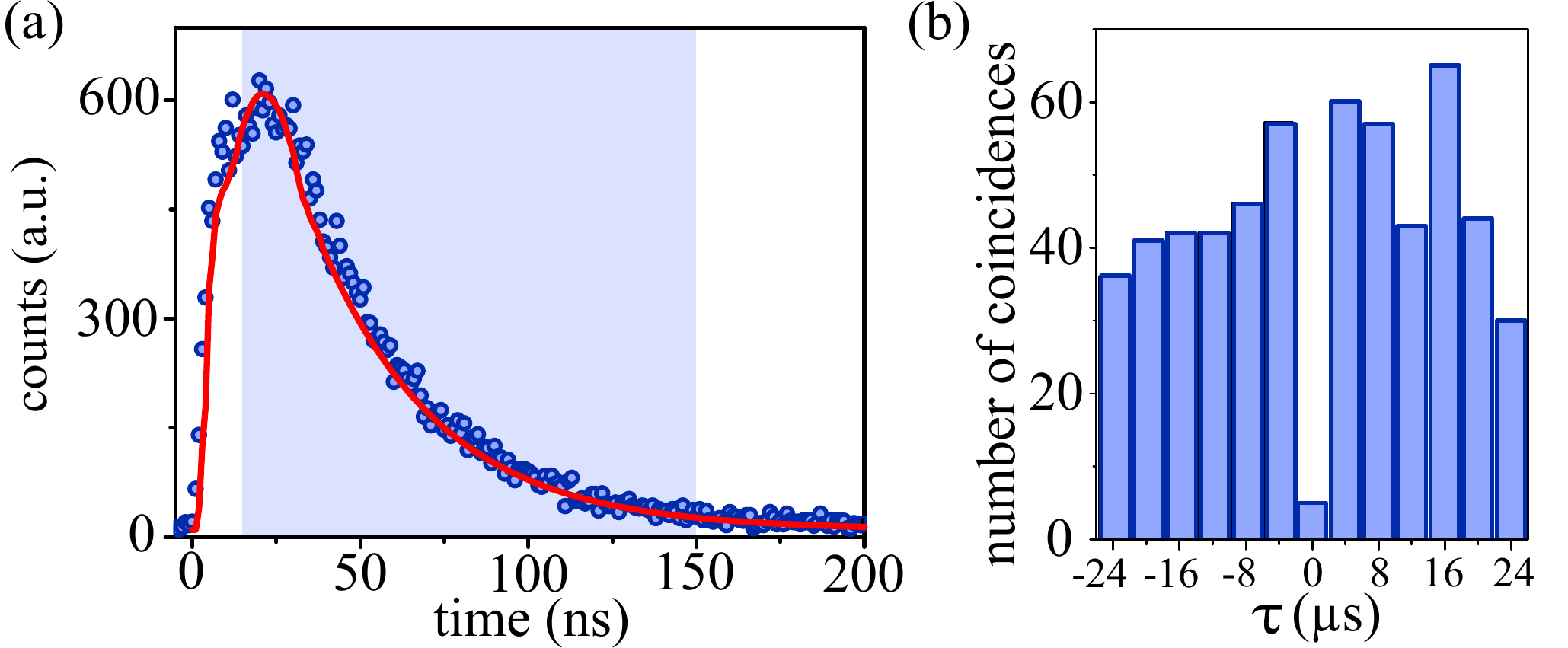}
\caption{(color online). \textbf{(a)} Time-arrival histogram. Solid line: Theoretical curve based on our cavity-QED parameters~\cite{PhysRevA.73.053807} and including background light. After $t=25\,$ns the intensity decays exponentially with a time constant $\tau_\text{hist}=35(1)\,$ns. \textbf{(b)} Histogram of the coincidence detection events of detector 1 and 2 separated by a time $\tau$. Only detection events within a 130~ns interval (indicated as blue shaded area in time-arrival histogram) are considered for the coincidence measurement and all detections events within this interval are summed.}
\label{fig:fig_2}
\end{figure}
 
We benchmark the prospects of using the polarization state of the extracted photons for faithful transfer of quantum information between ion and photon. To this end, we employ the  correlations between ion and photon produced during the emission process \cite{Blinov2004} (Figure 3a). First, the ion is prepared in the $m_J=-3/2$ ($\ket{\downarrow}$) Zeeman state of the $^2$D$_{3/2}$ manifold by optical pumping using $\pi$ and $\sigma^-$ polarized 935\,nm light in combination with 369\,nm light~\cite{Herskind:2009}. Next, a short pulse excites the ion to the $m_{J^\prime}=-1/2$ Zeeman state of the $^3$D[3/2]$_{1/2}$ manifold, and a photon is subsequently emitted into the cavity mode. Because we apply a magnetic field of approximately 3 Gauss along the cavity axis, only the $\sigma^-$ and $\sigma^+$ transitions are stimulated by the cavity. Consequently the emission of a $\sigma^+$ polarized photon leaves the ion in the $m_J=-3/2$ ($\ket{\downarrow}$) Zeeman state of the $^2$D$_{3/2}$ manifold whereas a $\sigma^-$ photon is accompanied by a decay to the $m_J=+1/2$ ($\ket{\uparrow}$) state. The small Zeeman splitting of the transition frequencies does not affect the coupling to the cavity because of the large cavity bandwidth. 

The overall state after the photon emission depends on the ratio of the Clebsch-Gordan coefficients of the $\sigma^+$ and $\sigma^-$ transitions. Thus, for the considered emission process the normalized $\ket{\text{ion}}\otimes\ket{\text{photon}}$ state $\ket{\Psi}$ reads   
\begin{equation} 
\ket{\Psi} =  \frac{\sqrt{3}}{2} \ket{\downarrow\text{, } \sigma^+} + \frac{1}{2}\ket{\uparrow\text{, } \sigma^-} .
\end{equation}
In order to measure the correlations between the spin state of the ion and the polarization of the photon, we read out the spin state of the ion after the photon emission. To this end, the $\ket{\uparrow}$  population is mapped onto the `bright' state, i.e. transferred to the $^2$S$_{1/2}$ state, whereas the $\ket{\downarrow}$ population remains in the `dark' $^2$D$_{3/2}$ state. Thereby subsequent fluorescence detection on the $^2$S$_{1/2}$-$^2$P$_{1/2}$ transition reveals the spin state.
\begin{figure}[!h]
\includegraphics[scale=0.48,angle=0]{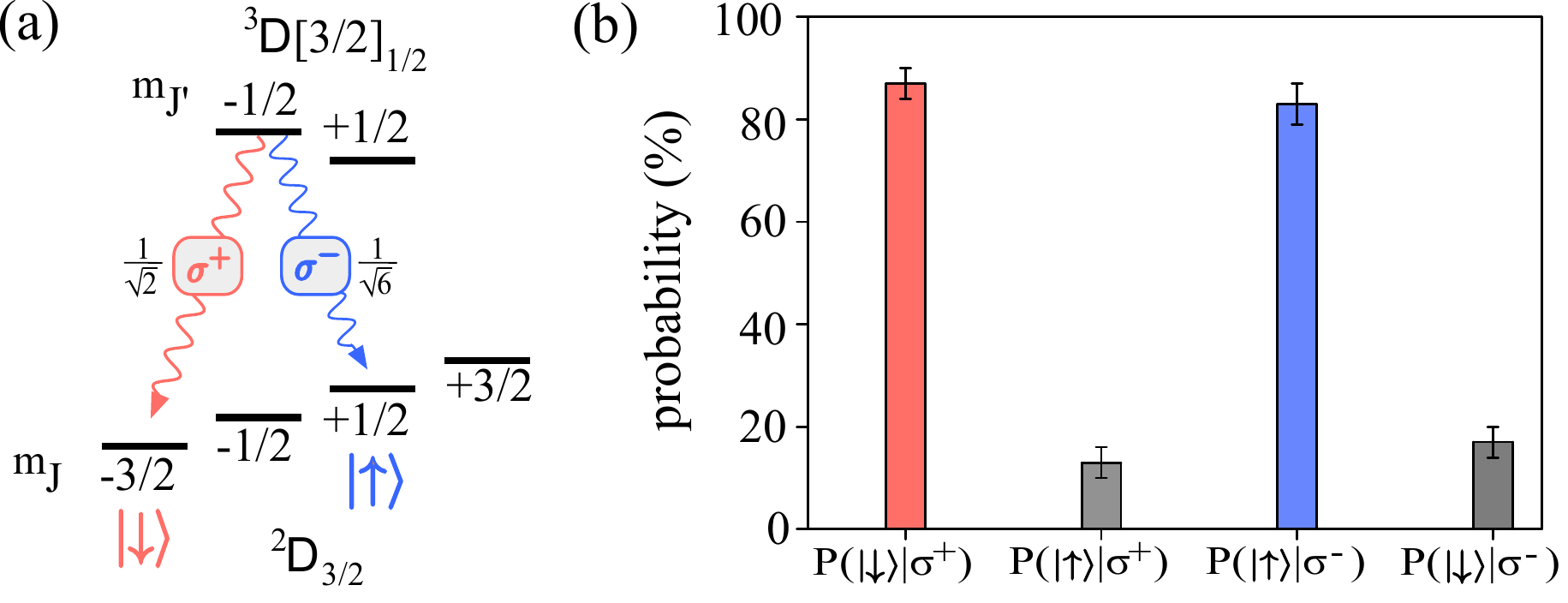}
\caption{\label{fig:fig_3}(color online). \textbf{(a)} Spin-photon correlations are created by the emission process subsequent to an excitation of the  $m_{J^\prime}=-1/2$  Zeeman level of the $^3$D[3/2]$_{1/2}$ state. The relevant Clebsch-Gordan coefficients are stated next to the transition arrows. \textbf{(b)} Spin state of the ion depending of the polarization of the detected photon.}
\end{figure}

Figure~3b shows the measured spin state conditioned on the detection of a $\sigma^+$ or a $\sigma^-$ polarized photon. We observe the expected strong correlation of $P\left(\ket{\downarrow}|~\sigma^+\right) = 87 (3)\%$ and $P\left(\ket{\uparrow}|~\sigma^-\right) = 83 (4)\%$. Furthermore the emitted light field contains more $\sigma^+$ than $\sigma^-$ polarized photons due to the unequal Clebsch-Gordan coefficients of the two transitions. The experimentally observed ratio of $\sigma^+$ and $\sigma^-$ polarized photons is  $r_{\sigma^+/\sigma^-}=2.1 (1)$. The deviations of the correlations and $r_{\sigma^+/\sigma^-}$ from the ideal state $\ket{\Psi}$ are in agreement with the expected errors from the state preparation, the background and the dark count rate of the SPCM. 

Finally, complementary to the photon emission experiments we study the absorption of the ion when the cavity mode is externally pumped by a faint laser beam \cite{PhysRevLett.98.193601,Specht2011}. In order to measure the  probability for incoming photons to excite the ion into the $^3$D[3/2]$_{1/2}$ state we take advantage of the large branching ratio of the decay from the $^3$D[3/2]$_{1/2}$ to the $^2$S$_{1/2}$ state. Thus, an excitation event will predominantly lead to a subsequent decay to the $^2$S$_{1/2}$ state which we read out by scattering on the 369\,nm transition. We therefore define the absorption per photon $p_\text{abs}$ as the probability for an incoming photon to transfer the ion from the $^2$D$_{3/2}$ to the $^2$S$_{1/2}$ state. 

The experimental sequence consists of preparing the ion in the $m_J=-3/2$ Zeeman level of the $^2$D$_{3/2}$ state by optical pumping, followed by the application of a $170\,\mu$s long probe pulse sent onto the HT-side of the cavity. The number of photons impinging upon the cavity is determined from the light reflected by the cavity. We vary the average number of photons reaching the cavity during the probe pulse between 30 and 80 and normalize the observed $^2$D$_{3/2}$-$^2$S$_{1/2}$ transfer probability to obtain the absorption probability per photon $p_\text{abs}$. 
\begin{figure}[h]
\includegraphics[width=0.45\textwidth,angle=0]{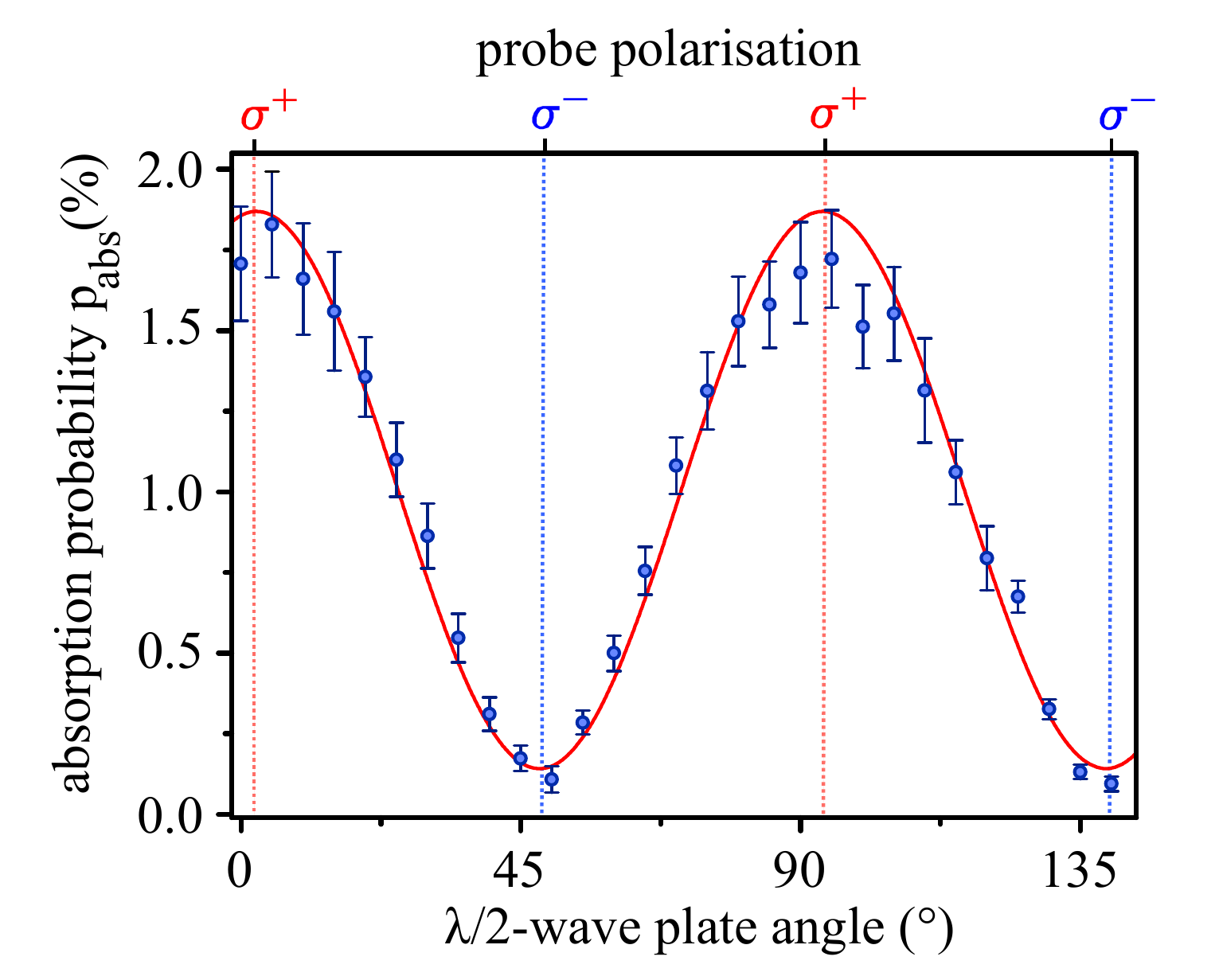}
\caption{\label{fig:fig_4}(color online). Absorption of a weak coherent pulse injected into the cavity mode normalized to the average number of photons in the probe pulse. The absorption probability $p_\text{abs}$ is the probability for a photon impinging upon the fiber-cavity to transfer the ion from the $^2$D$_{3/2}$ to the $^2$S$_{1/2}$ state. At $5^\circ $ and $95^\circ $ ($50^\circ $ and $140^\circ $) the polarization of the probe pulse is purely $\sigma^+$ ($\sigma^-$).}
\end{figure}

Figure~4 shows the expected oscillatory dependence of $p_\text{abs}$ on the polarization of the probe light with maxima and minima corresponding to $\sigma^+$ and $\sigma^-$ polarization, respectively. For $\sigma^+$ polarized light the absorption is $p_{\text{abs,}\sigma^+} = 1.8 (2) \%$. Compared to single ion experiments in free space \cite{PhysRevLett.105.153604} our observed absorption is almost 50\% larger despite the much more unfavourable level structure. Imperfect initialization of the spin state leads to a residual absorption of $\sigma^-$ polarized probe light with $p_{\text{abs,}\sigma^-} = 0.14 (2)\,\%$. The stated probabilities are a lower bound for the photon absorption because not every excitation event leads to a transfer to the  $^2$S$_{1/2}$ state. The natural fraction of $98.2\%$ is reduced by the Purcell effect and from the cavity-QED parameters we infer that the altered branching ratio is approximately 91\%. Thus, the absorption probabilities are slightly underestimated by 9\%. 

In the intermediate coupling regime ($g_{0}\approx \gamma<\kappa$) the absorption of coherent probe light is approximately given by the cavity-enhanced optical depth $2C_0$. However, the observable cooperativity in the absorption experiment is expected to be reduced compared to the aforementioned emission experiments by several factors: branching ratio (0.91), Clebsch-Gordan coefficient (0.75), state preparation (0.9), cavity in-coupling (0.8). After including these factors the measured cooperativity in absorption is 40(20)\% smaller than the cooperativity in emission. We attribute the residual discrepancy to uncompensated micromotion along the cavity axis. Micromotion induced sidebands lie within the cavity bandwidth but are resolved from the natural linewidth of the transition ($2\gamma < \Omega_\text{RF} < 2\kappa$) and therefore reduce the absorption of a weak near-monochromatic probe pulse stronger than the emission into the cavity mode.   

 
In summary, we have demonstrated efficient photon emission and absorption by the ion via the fiber cavity. We have achieved efficient coupling between cavity and fiber mode and showed that the distortion of the photon polarization due to the intrinsic single-mode fiber coupling is small. Further improvements of the efficiency can be made by employing cavities with higher finesse and improving the localization of the ion within the cavity mode. Recently it has been shown that absorption and scattering losses of high-reflective dielectric mirrors can be low enough to facilitate fiber-cavities with a finesse exceeding $10^5$~\cite{brandstaetter2013}. With these future technical improvements, the trap and cavity geometry presented here can realize not only a large cooperativity $C_0>1$ but also the strong coupling regime between a single ion and a single photon. 

We thank M. Atat{\"u}re for providing the amplitude modulator and Leander Hohmann for technical support. This work has been supported by the Alexander-von-Humboldt Professorship, EPSRC (EP/H005679/1), ERC (Grant No. 240335), ITN COMIQ and the  University of Cambrigde.

\newpage
\section{Supplementary material}
\subsection{Out-coupling efficiency of the cavity}
The probability of photons to leave the cavity through the high transmittance mirror is given by the ratio of the transmittance and the total sum of passive losses
\begin{equation}
\eta_\text{HT-mirror} = \frac{T_\text{HT}}{T_\text{HT}+T_\text{LT}+L} \approx 32(4)\%
\end{equation}
where $T_\text{HT}=100\,$ppm ($T_\text{LT}=10\,$ppm) is the transmittance of the high (low) transmittance mirror and $L=200 (50)\,$ppm is the sum of absorption and scattering losses per round-trip. 

In order to determine the mode matching between fiber and cavity mode the reflection of a resonant probe beam impinging upon the HT-side is measured. For perfect mode-matching the in-coupling on resonance is given by
\begin{equation}
\eta_\text{in} = 1 - \left(\frac{{T_\text{HT}-T_\text{LT}-L}}{T_\text{HT}+T_\text{LT}+L} \right)^2 \approx 90(5)\%~.
\end{equation}
The deviation of the experimentally observed in-coupling $\eta_\text{in,exp}=80(5)\%$ is caused by the mode mismatch $\epsilon$.
\begin{equation}
\epsilon = \frac{\eta_\text{in,exp}}{\eta_\text{in}} \approx 90(5)\%
\end{equation}

\subsection{Determination of $\bar{g}$}
As described in the main text we deduce the cooperativity $ C_0=0.032(8)$ from the number of photons emitted through the high-transmittance side of the cavity. Assuming an ideal-two level system ($ C_0= g_0^2/(2\kappa \gamma)$) this results in $g_{0\text{,obs}}=1.8(2)$. However, during the photon generation a bias magnetic field  of approximately 3 Gauss is applied along the cavity axis and therefore $\sigma^+$ and  $\sigma^-$ transitions are stimulated by the cavity field. Hence the observed emission rate is the sum of these two processes. From the observed cooperativity we calculate the coupling strength for the strongest transition  $\bar{g}_0$  by taking account of the Clebsch-Gordan coefficients.
\begin{equation}
\bar{g}_0 = \sqrt{\frac{3}{4}} g_{0\text{,obs}} = 1.6(2)
\end{equation}

\subsection{Optical pumping for state initialization}
For all spin dependent measurements presented in the main text, the ion is prepared in the $m_J=-3/2$ state of the $^2\text{D}_{3/2}$ manifold by optical pumping. In detail this is achieved by  driving the 369\,nm cooling transition and shining in a 935\,nm repump laser (`initialization laser'), where the polarization is adjusted such that it only couples $\sigma^-$ and $\pi$ transitions. Therefore the populations in the $m_J=\pm 1/2,\,+3/2$ states are transferred to the $^2\text{S}_{1/2}$-state, while the $m_J=-3/2$ population remains in the $^2\text{D}_{3/2}$ state. In entire preparation sequence takes $120\,\mu$s during which the 369\,nm light is continuously applied. During the first $100\,\mu$s the initialization laser is switched on for $1\,\mu$s every $10\,\mu$s and switched off completely after $100\,\mu$s. The 369\,nm light remains on for the residual $20\,\mu$s. With this sequence we achieve a state preparation in the $m_J=-3/2$ state of the $^2\text{D}_{3/2}$ manifold of approximately $90\%$.

\subsection{Spin-Photon correlation}
After a 935~nm photon has been detected at the HT-side of the cavity, the spin state of the ion is determined by mapping the population of one of the two relevant Zeeman substates onto the $^2\text{S}_{1/2}$ state which is read out via fluorescence detection on the $^2\text{S}_{1/2}$-$^2\text{P}_{1/2}$ transition. To achieve this mapping the ion is illuminated for about $2\,\mu$s with $\sigma^-$ and $\pi$ polarized light at 935~nm . If the ion is in the $m_J=-3/2$ state of the $^2\text{D}_{3/2}$ manifold it will not couple to this light field. Therefore it remains in the $^2\text{D}_{3/2}$ state and no fluorescence is detected when driving the $^2\text{S}_{1/2}$-$^2\text{P}_{1/2}$ transition.

\begin{figure}[!h]
\includegraphics[scale=0.35]{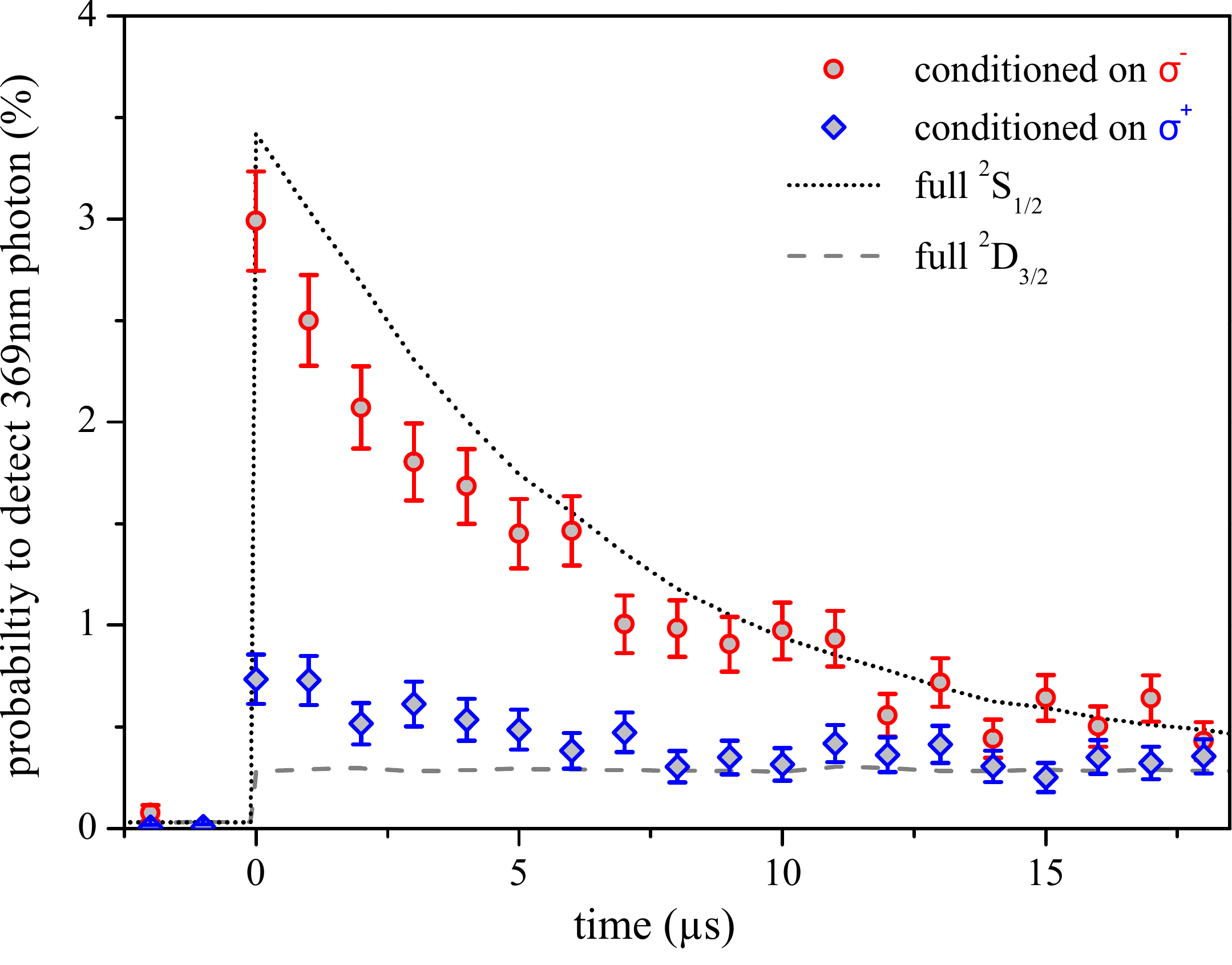}
\caption{\label{fig:supp_absorption}(color online). Read out of the $^2\text{S}_{1/2}$ state by detecting fluorescence on the $^2\text{S}_{1/2}$-$^2\text{P}_{1/2}$ transition at 369~nm. Dotted curve: the ion is initially prepared in the  $^2\text{S}_{1/2}$ state. Dashed curve:  the ion is initially prepared in the  $^2\text{D}_{3/2}$ state. Blue diamonds: fluorescence conditioned on prior detection of a 935~nm $\sigma^-$ polarized photon at HT-side of the cavity.  Red circles: fluorescence conditioned on prior detection of a 935~nm $\sigma^+$ polarized photon at HT-side of the cavity. All fluorescence curves decay exponentially to the background level because the $^2\text{S}_{1/2}$ state is depleted during the read out. The first $19\,\mu$s of each curve are integrated to obtain the conditional probabilities.   }
\end{figure}
If however the ion is left in $m_J=+1/2$ state and it will be pumped into the  $^2\text{S}_{1/2}$ state. Consequently fluorescence at 369\,nm is detected. In figure~suppl.~1 the fluorescence at the 369\,nm  conditioned on detection of a $\sigma^\pm$ is shown. Here we also show the calibration measurements of the 369\,nm fluorescence for an ion prepared in the $^2\text{S}_{1/2}$ state (bright state) and in the $^2\text{D}_{3/2}$ (dark state). The fluorescence decreases to its background level as the ion is pumped into the $^2\text{D}_{3/2}$ manifold during the read out. By summing up the probability of detecting a 369\,nm photon over the first $19\,\mu$s of the read out and compare this to the calibration measurements we find the conditional probabilities shown in figure~3 of the main text.

\subsection{Absorption Measurement}

\begin{figure*}[h!]
\includegraphics[scale=0.7]{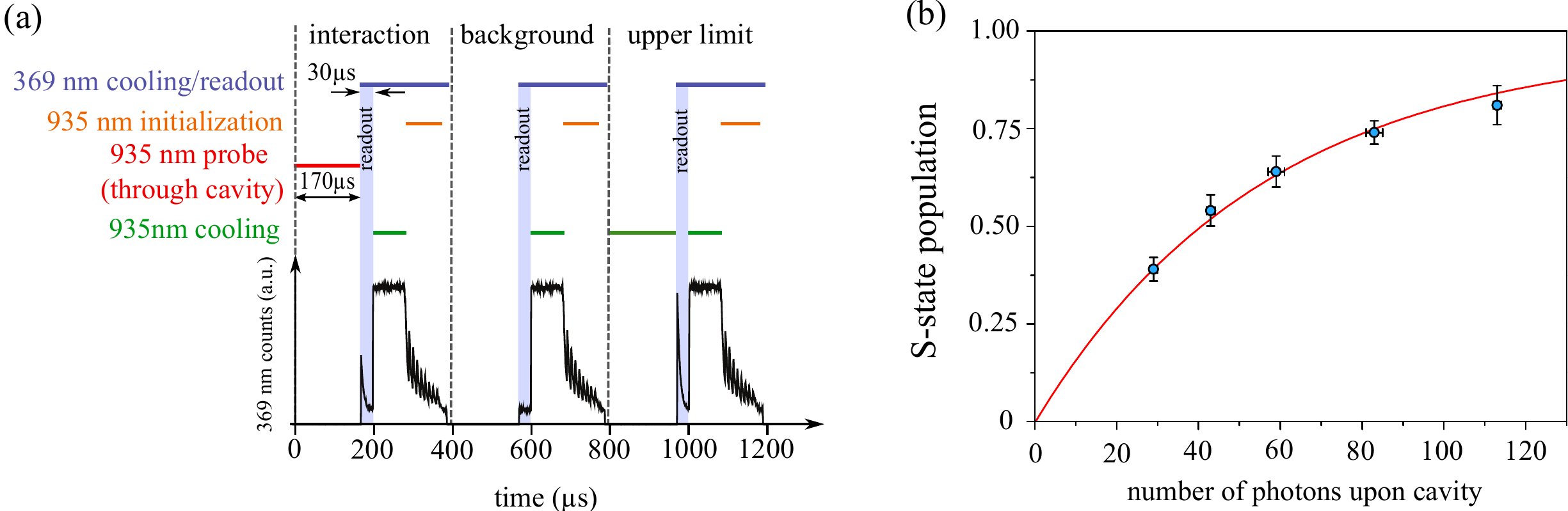}
\caption{\label{fig:supp_absorption}(color online). \textbf{(a)} Experimental sequence of absorption measurement. \textbf{(b)}  $^2\text{S}_{1/2}$ state population for repumping with $\sigma^+$ polarized light for different number of photons during the probe interval.}
\end{figure*}

In order to measure the absorption probability depending on the polarization of the probe light, we prepare the ion in the $m_J=-3/2$ state and measure the  $^2\text{S}_{1/2}$ state population via 369\,nm fluorescence after shining resonant probe light through the cavity for a duration of $170\,\mu$s (see figure~6a). We obtain calibration data for the fluorescence signal by repeating the measurement sequence without any probe light and as well as with the ion prepared in the $^2\text{S}_{1/2}$ state. By comparing the obtained 369\,nm fluorescence after the probe interval to the calibration data the probability to find the ion in the $^2\text{S}_{1/2}$ state $P_S$ is calculated.
By detecting the reflected light from the cavity, the number of photons impinging onto the cavity during the probe interval can be determined using the optical beam path efficiency. The  $^2\text{S}_{1/2}$ state population for probing with $\sigma^+$ light and different numbers of photons is shown in figure~6b. The curve is described by an exponential increase
\begin{equation}
P_S=1-e^{-n/n_{0}}
\end{equation}
where $n$ is the number of photons $n_0$ is the characteristic number of needed to increase the $^2\text{S}_{1/2}$ state population by $1/e$. The absorption probability per photon $p_{\text{abs}}$ is calculation from 
\begin{equation}
p_{\text{abs}}:=\frac{P_S}{n}=\frac{1}{n_{0}}.
\end{equation}
For the data shown in figure~6b we obtain $n_0=56(2)$ which results in $p_{\text{abs}}=1.8(2)\,\%$.

\end{document}